\documentclass[runningheads]{llncs}
\usepackage{threeparttable}
\usepackage{multirow}
\usepackage{graphicx}
\begin{document}

\title{On the Effectiveness of Neural Text Generation based Data Augmentation for Recognition of Morphologically Rich Speech}

\titlerunning{Effectiveness of Neural Text based Data Augmentation for ASR}

\author{Bal\'azs Tarj\'an\inst{1,2}
\and
Gy\"orgy Szasz\'ak\inst{1}
\and
Tibor Fegy\'o\inst{1,2}
\and
P\'eter Mihajlik\inst{1,3}
}

\authorrunning{B. Tarj\'an et al.}

\institute{Department of Telecommunications and Media Informatics, \\
Budapest University of Technology and Economics, Budapest, Hungary \\
\email{$\{$tarjanb,szaszak,mihajlik$\}$@tmit.bme.hu}
\and
SpeechTex Ltd., Budapest, Hungary \\
\email{fegyo@speechtex.com}
\and
THINKTech Research Center, V\'ac, Hungary
}

\maketitle              
\setcounter{footnote}{0}

\begin{abstract}

Advanced neural network models have penetrated Automatic Speech Recognition (ASR) in recent years, however, in language modeling many systems still rely on traditional Back-off N-gram Language Models (BNLM) partly or entirely.
The reason for this are the high cost and complexity of training and using neural language models, mostly possible by adding a second decoding pass (rescoring).
In our recent work we have significantly improved the online performance of a conversational speech transcription system by transferring knowledge from a Recurrent Neural Network Language Model (RNNLM) to the single pass BNLM with text generation based data augmentation.
In the present paper we analyze the amount of transferable knowledge and demonstrate that the neural augmented LM (RNN-BNLM) can help to capture almost 50\% of the knowledge of the RNNLM yet by dropping the second decoding pass and making the system real-time capable.
We also systematically compare word and subword LMs and show that subword-based neural text augmentation can be especially beneficial in under-resourced conditions.
In addition, we show that using the RNN-BNLM in the first pass followed by a neural second pass, offline ASR results can be even significantly improved.

\end{abstract}

\section{Introduction}

Deep learning has penetrated machine learning in the past years, including speech technology and language modeling in particular~\cite{Irie2019,sundermeyer2012lstm}.
Despite the success of this architectural paradigm shift, application of Neural Network Language Models (NNLM) in a single decoding pass is still challenging due to their structure and computational complexity.
NNLMs can still be used in ASR, when passing to the 2-pass decoding scheme: in the first pass, a small footprint generic Language Model (LM) is used, and the output of this step is a simplified recognition network with reduced search space.
On this reduced lattice, a second decoding pass is applied with the NNLM for rescoring the hypotheses obtained in the first pass.
Although by splitting the decoding into two parts we can leverage knowledge of the NNLMs and demonstrate significant Word Error Rate Reduction (WERR), it also introduces considerable processing delay~\cite{enarvi2017,Irie2019,sundermeyer2012lstm}.

Therefore, techniques exploiting the capabilities of NNLMs in a single-pass decoding approach have received particular attention recently~\cite{Singh2019,Suzuki2019}.
A possible technique is to augment the in-domain training data with a large text corpus generated by an NNLM~\cite{Adel2014,Deoras2011}.
Of course, there is a compromise: the augmented model is no more suitable for capturing long contexts, and lose capability to support continuous space features.
So far there has been no throughout evaluation of what NNLM capabilities can be transferred by neural text based data augmentation and how these compare to traditional Back-off N-gram Language Models (BNLM), especially for the morphologically rich languages.
The only exception is our earlier study for Hungarian~\cite{Tarjan2019a} showing that by combining subword lexical modeling with text based approximation of NNLM (referred to as RNN-BNLM) we can greatly improve the performance of an online ASR system.

In this paper we significantly extend our previous work:
(1) we quantify the amount of knowledge that can be transferred from the NNLM to single pass decoding with a BNLM augmented with data generated by the NNLM;
(2) we show that the performance of offline decoding can also be significantly improved if we apply the augmented model in the first-pass for generating the lattice;
(3) we evaluate the impact of training corpus size on the effectiveness of the data augmentation method.
Rich morphology, per se, results in extremely large vocabularies, which constitutes a challenge for language modeling.
Since data sparsity problems can be often handled by estimating language models on statically derived subword units (such as morphs)~\cite{Creutz2002,Kurimo2006487}, we will also evaluate morph-based models in our experiments.

In a related work, Suzuki et al.~\cite{Suzuki2019} use a domain balanced mixture of the training corpora to train a shallow RNNLM for text generation and improve speech recognition results for Japanese, Korean and English tasks.
For Korean subword-based language models are also utilized, but only for text generation, since in the language model of the ASR system subwords are merged.
Using subword units for language models and ASR has been mostly considered for Finnish and Estonian, which are morphologically very rich languages~\cite{Creutz2002,Kurimo2006487}.
In~\cite{enarvi2017}, the authors managed to outperform word-based baseline model on Finnish and Estonian conversations by training subword RNNLMs and utilizing them in the second pass to rescore ASR lattices.
N-gram based approximation of RNNLM was also investigated in a recent paper~\cite{Singh2019}, where subword and character-based models were trained for Finnish and Arabic OOV keyword search tasks.
Although the interpolation of approximated RNNLM and BNLM models improved OOV retrieval the proposed system was not evaluated on in-vocabulary tokens and no Word Error Rate (WER) was presented either.

\section{Data and methods}

\subsection{Database} \label{database}

\subsubsection{Conventional training data}

Data for modeling word units are taken from the Hungarian Call Center Speech Database (HCCSD).
The HCCSD corpus contains real conversations recorded in customer service centers. The conversations are transcribed and validated by human proofreaders.
A total of 3.4M word tokens could be used allowing for a dictionary of 100K distinct word forms.
In order to speed up training, the final vocabulary was limited to the most frequent 50K word forms.
The remaining Out-Of-Vocabulary (OOV) words were replaced with $\langle unk\rangle$ and the sentence endings were mapped to the $\langle eos\rangle$ symbol.
Training corpus statistics are summarized in Table~\ref{test_stat}.

\renewcommand{\arraystretch}{1.1}
\begin{table}[tbp]
\centering
\caption{Training and test database statistics}
\begin{tabular}{rccc}
\hline
                   & \textbf{~~~Training} & \textbf{~Validation~} & \textbf{~Evaluation~} \\ \hline
Duration {[}h:m{]} & ~~~290:07 & 7:31                & 12:12               \\
\# of word tokens       & ~~~3,401,775 & 45,773               & 66,312               \\
\# of morph tokens       & ~~~3,822,335 & 57,849               & 84,385               \\ \hline
\end{tabular}
\label{test_stat}
\end{table}

\subsubsection{Morph segmented training data}

Morphologically rich languages like Hungarian show heavy agglutination and hence vocabulary gets much larger.
This also results in higher variability regarding word sequences, and estimation of model parameters becomes less accurate.
Segmenting words into smaller units is driven by the idea to both decrease vocabulary size and increase sequential consistency in morph sequences~\cite{Kurimo2006487}.
Morfessor~\cite{Creutz2002} is a popular algorithm for segmenting words into subword units as it iteratively finds the optimal decomposition of vocabulary words into subword units, called \textit{morphs}.
In~\cite{Smit} it was shown that Morfessor can outperform the nowadays so popular character-level Byte Pair Encoding (BPE) algorithm.

The training corpus contained 3.8M units after applying Morfessor and decomposing words into morphs (see Table~\ref{test_stat}).
The number of vocabulary entries decreased to around 1/3 of the word vocabulary, that is to 32K entries covering the same text corpora as the word based model.
The morph vocabulary was finally limited to 30K morphs based on frequency, in order to provide enough training samples to $\langle unk\rangle$.
Morphs in non-word-initial position were additionally tagged by the `+' sign to preserve this syntactic information relative to original word boundaries.
The following example illustrates a morph-based tokenization (decomposition) of the sentence ‘well I will discuss this with my wife’:\\
\\
Conventional tokenization: \textit{h\'at megbesz\'elem a nejemmel}\\
Morph-based tokenization: \textit{h\'at meg +besz\'el +em a nejem +mel}

\subsubsection{Development and test data} \label{test_data}

For validation and testing, two further disjoint data sets were created using 20 hours of conversations, reserved from the HCCSD corpus (see Table~\ref{test_stat}).
The validation set is required for the optimization of the hyperparameters (e.g. Morfessor segmentation, control training of language models), whereas the evaluation set is used for performance evaluation and comparison of the models.

\subsection{Language modeling methods} \label{lm}

\subsubsection{Back-off n-gram models}

N-gram models are statistical, count-based models estimated on large text corpora. Back-off N-gram Language Models (BNLM) formed the state-of-the-art in language modeling for ASR over several decades, and still today, for a number of tasks they are the primary choice, especially in systems requiring real-time or smaller footprint setups.
All BNLMs in this work are estimated with the SRI language modeling toolkit~\cite{Stolcke2002} and smoothed with Chen and Goodman's modified Kneser-Ney discounting.

\subsubsection{Recurrent neural language model} \label{rnnlm}

We implemented\footnote{https://github.com/btarjan/stateful-LSTM-LM} a 2-layered LSTM structure according to the scheme presented in~\cite{Tarjan2019a}.
After fine-tuning the hyperparameters on the validation set, we use a batch size of 32 sequences, composed of 35 tokens each (tokens can be either words or morphs).
LSTM states are preserved between the batches (stateful LSTM).
The 650 dimensional embedding vectors were trained from scratch, as transfer learning from existing Hungarian pretrained embeddings proved to be suboptimal.
After trying several optimizers, we decided on the traditional, momentum accelerated Stochastic Gradient Descent (SGD) algorithm.
The initial learning rate was set to 1, which is halved after every epoch where the cross entropy loss increases.
To prevent overfitting dropout layers are used with keep probabilities of 0.5.
Early stopping with a patience of 3 epochs is also applied.

\subsubsection{Text generation based data augmentation} \label{sec:approx}

Approximation of a NNLM with a back-off ngram language model can be achieved in several different ways~\cite{Adel2014,Deoras2011}.
In~\cite{Adel2014} three such methods are described and evaluated, coming to a conclusion that the so called text generation based data augmentation yields the best results.
The main idea of this approach is to estimate the BNLM parameters from a large text corpus generated by a NNLM.
In our work, we generated 100 million words/morphs with the corresponding RNNLM (RNN-BNLM 100M) that was formerly trained on the in-domain training set.
In order to get an insight how the corpus size influences the language model capabilities, we also generated a larger text corpus with 1 billion morphs (RNN-BNLM 1B).
To achieve the best results the models trained on augmented text (RNN-BNLMs) are interpolated with the baseline models (BNLM + RNN-BNLM).
Interpolation weights are optimized on the development set.

\section{Results and Discussion} \label{results}

\subsection{Experimental setup}

40 dimensional MFCC vectors were used as input features for a Factored Time Delay Neural Network (TDNN-F) acoustic model trained applying LF-MMI criterion in a similar manner as in~\cite{Povey2018} using the Kaldi Toolkit~\cite{Povey:192584}.
The matrix size (hidden-layer dimension) was 768 and the linear bottleneck dimension was 80 resulting in a total number of 6M parameters in the twelve hidden layers.
Acoustic and language model resources were compiled into weighted finite-state transducers and decoded with our in-house ASR decoder, called VoXerver.

\begin{table}[tbp]
\centering
\caption{WER of the online ASR system using the proposed language models}
\begin{tabular}{clccccc}
\hline
\textbf{\begin{tabular}[c]{@{}c@{}}Token~\\type~\end{tabular}} & \textbf{Model}                      & \textbf{\begin{tabular}[c]{@{}c@{}}\# of\\n-grams\\{[}million{]}\end{tabular}} & \textbf{\begin{tabular}[c]{@{}c@{}}Memory\\usage\\{[}GB{]}\end{tabular}} & \textbf{\begin{tabular}[c]{@{}c@{}}WER~\\{[}\%{]}~\end{tabular}} & \multicolumn{2}{c}{\textbf{\begin{tabular}[c]{@{}c@{}}WERR over\\Word/Morph\\BNLM {[}\%{]}\end{tabular}}} \\ \hline
\multirow{3}{*}{\rotatebox[origin=c]{90}{Word}}                                         & BNLM                                & 5.0                                                                              & 1.3                                                                        & 21.9                                                            &                                                                                                     &  \\
                                                              & RNN-BNLM 100M                       & 4.8                                                                              & 0.9                                                                        & 22.5                                                            & ~~-2.6*                                                                                                 &  \\ \cline{2-7} 
                                                              & BNLM + RNN-BNLM 100M                & 7.0                                                                              & 1.5                                                                      & 21.3                                                            & ~~~2.7*                                                                                               &  \\ \cline{2-7} \hline
\multirow{6}{*}{\rotatebox[origin=c]{90}{Morph}}                                        & BNLM                                & 5.1                                                                              & 1.0                                                                        & 21.1                                                            & ~~~3.4*                                                                                                &  \\
                                                              & RNN-BNLM 100M                       & 8.5                                                                              & 1.1                                                                        & 21.1                                                            & ~~~3.7*                                                                                                & 0.3~~\\
                                                              & RNN-BNLM 1B                         & 7.2                                                                              & 0.9                                                                        & 20.5                                                            & ~~~6.4*                                                                                                & 3.2*\\ \cline{2-7} 
                                                              & BNLM + RNN-BNLM 100M                & 7.9                                                                              & 1.1                                                                      & 20.4                                                            & ~~~6.8*                                                                                                & 3.5*\\ \cline{2-7}
                                                              & \multirow{2}{*}{BNLM + RNN-BNLM 1B}   & 7.2                                                                              & 1.1                                                                      & 20.2                                                            & ~~~7.7*                                                                                                & 4.5*\\
                                                              &                                     & 46.6                                                                             & 5.9                                                                        & 19.9                                                            & ~~~8.8*                                                                                                & 5.6*\\ \cline{2-7} \hline
\end{tabular}
\begin{tablenotes}
\item * sign indicates significant difference compared to Word or Morph-based BNLM models and was tested with Wilcoxon signed-rank test (p \textless~0.05).
\end{tablenotes}
\label{wer}
\end{table}

\subsection{Online ASR results with data augmentation}

We perform single-pass decoding with 4-gram BNLM and RNN-BNLM models and calculate WER on the evaluation set.
In order to ensure the fair comparison among the modeling approaches, we pruned each RNN-BNLM so that they had similar runtime memory footprint as the baseline BNLM models ($\sim$1 GB).
The most promising model, where the baseline is augmented with 1 billion token corpus (BNLM + RNN-BNLM 1B) however, is also evaluated in a setup allowing for larger memory consumption to determine the full capability of the model.

\subsubsection{Results with original training corpus}

First we discuss the online ASR results (see Table~\ref{wer}) obtained with models trained on the original training corpus (3.4M word/3.8M morph tokens).
The n-gram model estimated on the corpus that was generated with the word-based RNNLM (RNN-BNLM 100M) has a slightly higher WER than the baseline word-based BNLM (2.6\% relative WER increase), but with the interpolated model (BNLM + RNN-BNLM 100M) we are able to significantly outperform both of them (2.7\% rel. WERR).

Switching to subword setups we observed the following results: the simple act of replacing words with subwords in the baseline BNLM already yields a significant WER reduction (3.4\% rel.).
The LM trained on the 100-million-morph generated corpus (RNN-BNLM 100M) has the same WER as the morph-based BNLM (21.1\% WER).
However using a ten times larger corpus to train the approximative model reverses this trend: morph-based RNN-BNLM 1B model is the first augmentation model that outperforms a baseline BNLM by itself, without taking any benefit from interpolation (20.5\% WER).
When adding interpolation, we can leverage a further increase in performance.
BNLM + RNN-BNLM 1B model can reduce WER of morph-based BNLM by 5\% or even 6\% if runtime memory consumption is not a restricting factor.
All in all, with morph-based neural text generation we managed to reduce the WER of our call center speech transcription system by 9\% relative while preserving real-time operation.

\begin{figure}[tbp]
\centering
\includegraphics[width=1.0\textwidth]{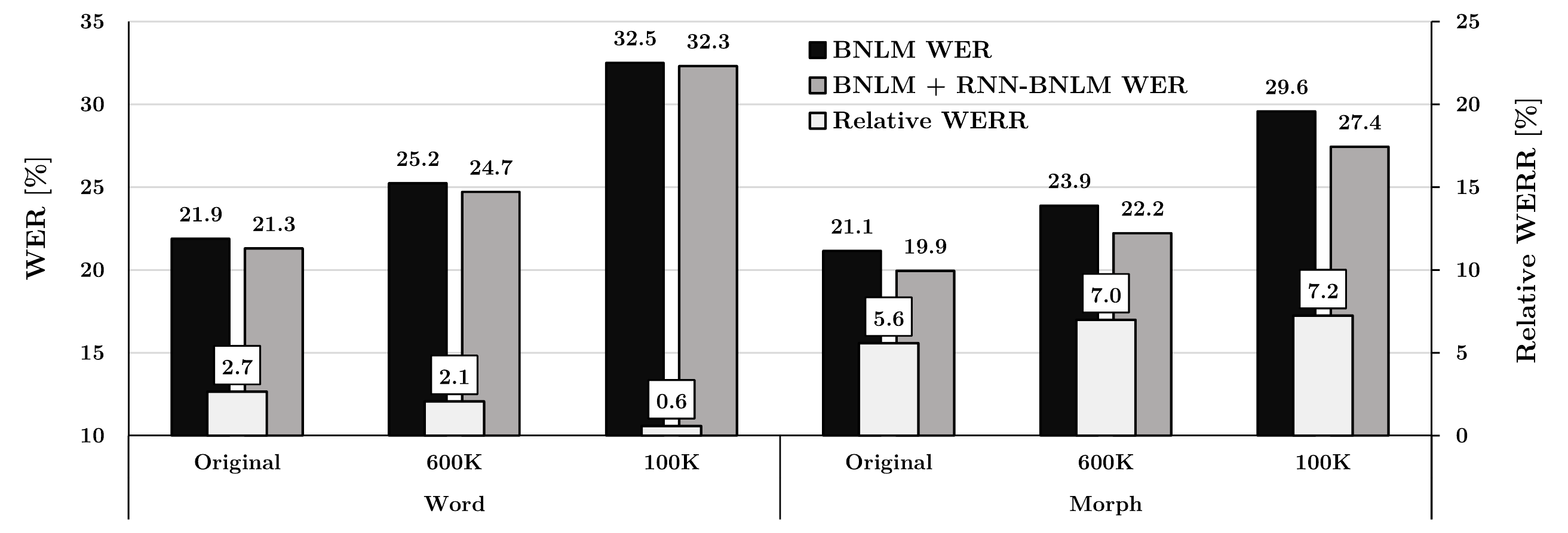}
\caption{Impact of training data limitation on the WER of baseline and augmented LMs and the corresponding relative WERRs}
\label{fig:size_WER-WERR}
\end{figure}

\subsubsection{Impact of training corpus size}
The RNNLM used to generate augmentation data is trained on in-domain training corpus, hence we assume that the amount of available training data is closely related to the effectiveness of this modeling approach.
In order to confirm this hypotheses, we repeated the experiments from the previous section, but we limited the size of training database (see Fig.~\ref{fig:size_WER-WERR}).
The original corpus containing 3.4M tokens was reduced to two smaller corpora following a log-uniformly spaced scale (600K and 100K).
We found that in case of word-based modeling the less training data is used the smaller is the benefit of data augmentation.
In contrast, morph-based augmented models even increase their advantage over the baseline for smaller training sets.
Based on the above, we conclude that text based augmentation can be indeed effective in under-resourced conditions, if it is paired with subword lexical modeling approach so that the RNNLM has enough samples for learning.

\subsection{Comparing online and offline ASR results}

\begin{figure}[tbp]
\centering
\includegraphics[width=1.0\textwidth]{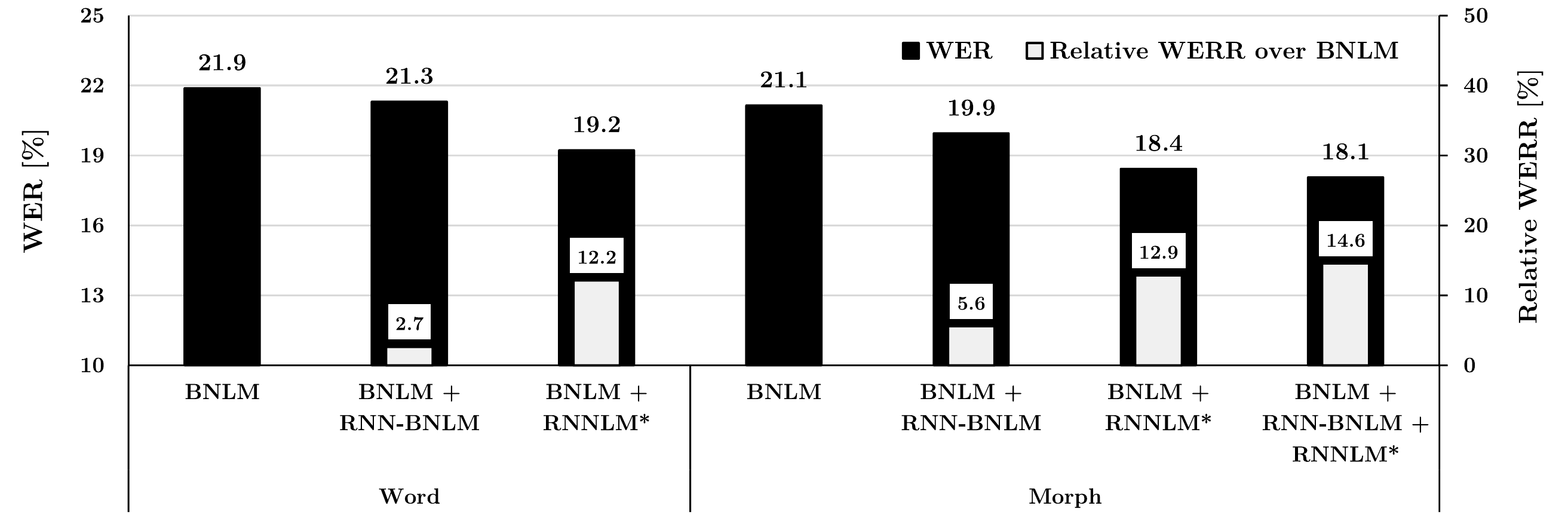}
\caption{A comparison of online (baseline LM: BNLM, augmented model: BNLM+RNN-BNLM) and offline (rescoring BNLM with RNNLM: BNLM+RNNLM, rescoring BNLM+RNN-BNLM with RNNLM: BNLM+RNN-BNLM+RNNLM) ASR results with word and morph based lexical modeling. * indicates offline, 2-pass decoding}
\label{fig:online-offline_WER-WERR}
\end{figure}

In this section we compare the performance of the original RNNLM applied for 2-pass, offline decoding and the RNN-BNLMs in order to assess the amount of knowledge that can be transferred to the online ASR system (see Fig.~\ref{fig:online-offline_WER-WERR}).
With offline, 2-pass decoding, the baseline WER can be reduced by $\sim$12-13\% (BNLM + RNNLM).
Word-based augmentation can capture 22\% of this WERR as it reduces the WER by 2.7\% compared to the 12.2\% of 2-pass decoding.
Using morph-based lexical modeling and a 10 times larger augmentation corpus the relative WERR can be increased to 5.6\% (Morph BNLM + RNN-BNLM).
On this basis we can conclude that up to 45\% of the WERR (5.6\% from 12.9\%) potential hold by the RNNLM can be transferred to the first pass of the decoding.

Text based data augmentation was introduced to improve online ASR results by transferring knowledge from the neural model to the BNLM.
However, we found that even offline speech recognition can benefit from this approach.
The last column in Fig.~\ref{fig:online-offline_WER-WERR} shows that significant (p=0.01) WERR can be achieved if the lattice used for rescoring is generated with the augmented model (BNLM + RNN-BNLM + RNNLM) instead of the original BNLM (BNLM + RNNLM).

\section{Conclusions} \label{conclusions}

In this paper neural LMs were used to transfer their knowledge to traditional back-off LMs by generating samples for probability estimation.
The morphological complexity of Hungarian was treated by using morph-based models evaluated on a call center ASR task.
We found that by generating a text with 1 billion morphs, the WER can be reduced by 9\% relative while preserving real-time operation.
The investigated neural text based data augmentation technique proved to be especially effective in under-resourced conditions provided that subword-based modeling is applied.
With the augmented LMs we managed to transfer $\sim$45\% of WERR of the offline, 2-pass configuration to our online system.
Finally, we also showed that augmented LMs can improve not only online but offline ASR results if they are used for generating the lattice for the 2nd decoding pass.

\section*{Acknowledgements}
The research was supported by the CAMEP (2018-2.1.3-EUREKA-2018-00014) and NKFIH FK-124413 projects.

\bibliographystyle{splncs04}
\bibliography{tsd1054}

\end{document}